\documentclass[11pt]{article}
\usepackage[english]{babel}
\usepackage[T1]{fontenc}
\usepackage[normalem]{ulem}
\usepackage{xcolor} 
\usepackage{amssymb, amsfonts, amsmath, amsthm} 
\usepackage{stmaryrd} 
\usepackage{braket} 
\usepackage{physics} 
\usepackage{enumerate} 
\usepackage{caption} 
\usepackage{tikz} 
\usetikzlibrary{chains, positioning, shapes.symbols,shapes,arrows}
\usetikzlibrary{angles,quotes}
\usetikzlibrary{graphs,graphs.standard,quotes}
\usepackage{pgfplots}
\pgfplotsset{compat=1.16}
\usepackage{graphicx}
\usepackage{subfiles} 
\usepackage{hyperref}
\usepackage[ruled,vlined]{algorithm2e}
\usepackage{circuitikz}
\usepackage{multicol} 
\usepackage[nottoc]{tocbibind} 
\usepackage[utf8]{inputenc}
\usepackage[top=1 in,bottom=1in, left=1 in, right=1 in]{geometry}
\newtheorem{theorem}{Theorem}
\newtheorem{proposition}{Proposition}
\newtheorem{lemma}{Lemma}
\newtheorem{corollary}{Corollary}

\newtheorem{remark}{Remark}
\usetikzlibrary{shadings}
\usetikzlibrary{arrows.meta}

\begin{document}

\title{Unravelling mean-field Lindblad equation}
\author{Sofiane Chalal \thanks{L2S, CentraleSupélec, Université Paris-Saclay. \texttt{sofiane.chalal@centralesupelec.fr}}
\and Nina H. Amini \thanks{CNRS. L2S, CentraleSupélec, Université Paris-Saclay. \texttt{nina.amini@centralesupelec.fr}}}

\maketitle

\begin{abstract}
{\small \bf We propose a mean-field particle Monte Carlo method for simulating the 
$N$-body Lindblad equation. We provide a convergence result showing that a system of interacting particles converges to the corresponding nonlinear Lindblad equation in the large-$N$ limit.}
\end{abstract}

\section*{Introduction}
Open quantum systems are systems that interact with an external environment, leading to dissipation (decoherence). Under certain regimes called the Born–Markov approximation, we can describe the system's evolution by averaging over the environment. The state of the system then evolves according to the quantum master equation (also known as the Lindblad or GKLS equation); see \cite{GKLS17History} for a historical review of its derivation.  

Unlike a closed quantum system, where the state is represented by a wavefunction (i.e., a vector in a Hilbert space), the interaction with the environment that induces dissipation causes the state to lose its purity. As a result, it becomes mixed, and is therefore described by a density matrix.

Nevertheless, if we consider the environment as performing a continuous measurement, we can describe the state’s evolution as a stochastic evolution of a pure state. By taking the average of these pure-state evolutions, we recover the dynamics described by the master equation. This is the idea behind quantum trajectories (see Chapters 3 and 4 \cite{wiseman09quantum}). 

A simulation of the quantum master equation using system of quantum trajectories can be performed instead of solving it for the density matrix. An important advantage of the quantum trajectory approach—first highlighted by Dalibard, Castin, and Mølmer \cite{dalibard92wave},\cite{molmer93monte} by Gisin and Perceival in \cite{gisin92quantum} and by Carmichael \cite{carmichael09open}—is that one only needs to handle a vector of dimension $d$, as opposed to a density matrix with $d^2$ elements. Hence, simulations become feasible in situations where directly solving the master equation would be impractical due to the large dimension of the system’s state space. In many applications, one is primarily interested in the steady-state density matrix. In such cases, it is often more convenient to replace ensemble averages by a time average taken over a single trajectory. Furthermore, the simulation approach is well-suited to execution on a distributed network of computers, which significantly increases computational power. Because the different wave-function realizations are statistically independent, the algorithm can be parallelized in a straightforward manner.

The mean-field approach has been widely used to deduce simplified models representing large number of interacting systems \cite{chaintron22I}, \cite{chaintron22II}. In quantum setting  Spohn \cite{spohn80kinetic} first established the foundational principle of the law of large numbers for a system of interacting quantum particles, yielding what is now known as the non-linear Schrödinger (or Hartree) equation. For open quantum systems, mean-field theory has been studied extensively in several works; see, for example, \cite{breuer07theoryopen}, \cite{prataviera14many}, \cite{merkil12} and the references therein. Motivated by building a quantum analogue to mean-field game theory, Kolokoltsov rigorously derived the mean-field equation for interacting quantum particles subject to continuous measurements \cite{kolokoltsov21law}, \cite{kolokoltsov21qmfgcounting}, \cite{kolokoltsov22qmfg}. This leads to a new non-linear stochastic Schrödinger-Hartree equation, which is a stochastic differential equation of McKean–Vlasov type.

In this letter, we use the equation derived in \cite{kolokoltsov22qmfg} to justify the mean-field approximation of the $N$-body Lindblad equation. Specifically, we employ a particle system to approximate the McKean–Vlasov dynamics. By averaging over the interacting trajectories, we obtain an approximation to the nonlinear Lindblad dynamics.

\subsection*{Notation}
The sets of natural, and complex numbers are denoted  $\mathbb{N},\mathbb{C}$, we denote $\mathfrak{R}(z)$ the real part of any complex number $z \in \mathbb{C}$. We denote $\mathcal{M}_{d}(\mathbb{C})$ set of complex square matrices of size $d \in \mathbb{N}$. We denote with roman symbol the imaginary number $\mathrm{i} := \sqrt{-1}.$

We fix  $\mathbb{H}$ notation for Hilbert space. We shall use both the notation $\psi$ and the "ket" $\ket{\psi}$ for a vector in $\mathbb{H}$, and $\bra{\psi}
$ the "bra" for an element of the dual space $\mathbb{H}^{\star} \simeq \mathbb{H}$.

The set $\mathcal{B}(\mathbb{H})$ be the set of linear bounded operator on $\mathbb{H}$. The set $\mathcal{B}_{1}(\mathbb{H})$ of trace-class and $\mathcal{B}_{2}(\mathbb{H})$ the set of Hilbert-Schmidt operators. For every $ {O} \in \mathcal{B}(\mathbb{H})$, denote by ${{O}}^{\dag}$ its adjoint operator. 
For any   ${O}_{a},{O}_{b} \in \mathcal{B}(\mathbb{H})$, set $[{O}_{a},{O}_{b}] := {O}_{a}{O}_{b} - {O}_{b}{O}_{a}$  and $\{{O}_{a},{O}_{b}\} := {O}_{a}{O}_{b} + {O}_{b}{O}_{a}$.

For every operator $O \in \mathcal{B}(\mathbb{H})$ set $\langle O\rangle_{\psi} := \bra{\psi}O\ket{\psi}$. For any operator ${O} \in \mathcal{B}(\mathbb{H})$ and for $l \in \{1,\dots,N\}$ denote 
$\mathbf{O}_{l} := \mathbf{1}\otimes\dots\otimes{O}\otimes\dots\otimes\mathbf{1}$ the operator on $\mathcal{B}_{1}(\mathbb{H}^{\otimes N})$ that acts only on the $l$-th Hilbert space.

For any operator $B \in \mathcal{B}(\mathbb{H}\otimes\mathbb{H}) $ and for $l,l' \in \{1,\dots,N \}$ denote $\mathbf{B}_{ll'}$ the operator on $\mathcal{B}(\mathbb{H}^{\otimes N})$ that acts only on $\mathbb{H}_{l}$ and $\mathbb{H}_{l'}$.

\section*{Setting}
The starting point is to fix a configuration space  $(\mathfrak{X},\Omega_{\mathfrak{X}},\mu)$ that form measure space and from which construct the Hilbert space of one particle\footnote{It's not really a restriction to proceed in this way,  complex separable Hilbert has isomorphism $\mathbb{H} \simeq l^2_{\mathbb{C}}(\mathbb{N}) \simeq L^{2}_{\mathbb{C}}(\mathbb{N},\lambda)$ with $\lambda$ a counting measure.} 
$$ \mathbb{H} := L^{2}_{\mathbb{C}}(\mathfrak{X},\mu)$$
We consider $N$ identical quantum particles\footnote{More precisely $N$ identicals bosons.} associated with a certain sites $ l = 1,2,\dots, N$ of a lattice\footnote{A spin lattice for example }.The Hilbert space of the $l-$th particle is denoted by $\mathbb{H}_{l}$, then total Hilbert space of the whole system is given by tensor product, 
\begin{align*}\mathbb{H}^{N} &:= \bigotimes_{l=1}^{N}\mathbb{H}_{l},\\
            &= L^{2}_{\mathbb{C}}(\mathfrak{X}^{N},\mu^{\otimes N})
\end{align*}

A state of system is a positive trace-class, self-adjoint operator of trace $1$, i.e $\boldsymbol{\bar{\rho}}^{N} \in \mathcal{S}(\mathbb{H}^{N})$ where,
$$ \mathcal{S}(\mathbb{H}^{N}) := \Big\{\boldsymbol{\rho} \in \mathcal{B}_{1}(\mathbb{H}^{N}) \;\;|\:\; \boldsymbol{\rho}^{\dagger} = \boldsymbol{\rho},\; \boldsymbol{\rho}\geq 0,\; \tr(\boldsymbol{\rho})=1\Big\}$$

We say that the system is in pure state $\boldsymbol{\rho}$ if one of the following equivalent statement holds :
\begin{enumerate}
    \item There exist $\psi \in \mathbb{H}$ such that $\boldsymbol{\rho} = \ket{\boldsymbol{\psi}}\bra{\boldsymbol{\psi}}$,
    \item $\tr(\boldsymbol{\rho}^2) = 1$,
    \item $\text{rank}(\boldsymbol{\rho}) = 1$.
\end{enumerate}

We attach self-adjoint free hamiltonian $H^{\dagger} = H \in \mathcal{B}(\mathbb{H})$ for each particle and we consider a symmetric pairwise interaction between two particles represented by a Hilbert-Schmidt operator $ A \in \mathbb{B}_{2}(\mathbb{H}\otimes\mathbb{H})$. This operator $A$ can be viewed as an element in  $ L^2_{\mathbb{C}}(\mathfrak{X}^{2},\mu^{\otimes 2})$  with kernel $a$ with the following proprieties :
{\small \begin{align*}
\|a\|^{2}_{2} &= \int_{\mathfrak{X}^4}|a(x,y;x',y')|^{2}\mu(\mathrm{d}x\mathrm{d}y\mathrm{d}x'\mathrm{d}y') < \infty\\
a(x,y;x',y') &= a(y,x;y',x'), \quad a(x,y;x',y') = \overline{a(x',y';x,y)}
\end{align*}
\begin{align*} 
\quad &A : L^2_{\mathbb{C}}(\mathfrak{X}\times\mathfrak{X},\mu^{\otimes 2}) \to
L^2_{\mathbb{C}}(\mathfrak{X}\times{\mathfrak{X}},\mu^{\otimes 2})\\
  Af((x,y)) &:= \int_{\mathfrak{X}^2}a(x,y;x',y')f(x',y')\mu(\mathrm{d}x'\mathrm{d}y')
\end{align*}}

The total Hamiltonian of the system is then given by

{\small$$ \mathbf{H}^{N} = \sum_{l=1}^{N} \tilde{\mathbf{H}}_{l} + \frac{1}{N}\sum_{l > l'}^{N}\mathbf{A}_{ll'}$$}

Each particle interacts with the environment via quantum channel $L \in \mathcal{B}(\mathbb{H})$, then tracing out the environment and taking the Born-Markov approximation one obtains the resulting $N$-body dynamics, which is governed by a Lindblad equation\footnote{Also known as the GKLS equation or the quantum master equation.}.

{\small \begin{align}\label{NLindblad}
\frac{\mathrm{d}\boldsymbol{\bar{\rho}}_t^{N}}{\mathrm{d}t} &= -\mathrm{i}{[\mathbf{H}^{N},\boldsymbol{\bar{\rho}}_{t}^{N}]}
+ \sum_{l=1}^{N} \Big({\bf L}_l\boldsymbol{\bar{\rho}}_{t}^{N}{\bf L}^{\dagger}_{l} - \frac{1}{2}\big\{{{{\bf L}^{\dagger}_{l}}}{\bf L}_{l},\boldsymbol{\bar{\rho}}_{t}^{N}\big\}\Big)\\
\boldsymbol{\bar{\rho}}_{0}^{N} &= \bigotimes_{{l=1}}^{N}\rho_{0}, \;\; \rho_{0} \in \mathcal{S}(\mathbb{H}^N).\nonumber
\end{align}}

One way to view Lindbladian dynamics is to say that the environment continuously performs a measurement on the system, but this measurement is not read out. Therefore, if one takes the measurement into account, the system follows a stochastic trajectory in phase space, and its dynamics is described by what is called a quantum trajectory \footnote{or Belavkin equation}.  

Then the $N$-body Quantum trajectories takes the following form :  
\begin{itemize}
\item Normalize - \textbf{Non}linear
\small{\begin{align}\label{Nnormalize}
\mathrm{d}\boldsymbol{\rho}_t^{N} &= -\mathrm{i}{[\mathbf{H}^{N},\boldsymbol{\rho}_{t}^{N}]}\mathrm{d}t
+ \sum_{l=1}^{N} \Big({\bf L}_l\boldsymbol{\rho}_{t}^{N}{{\bf L}^{\dagger}_{l}} - \frac{1}{2}\big\{{{\bf L}^{\dagger}_{l}}{\bf L}_{l},\boldsymbol{\rho}_{t}^{N}\big\}\Big)\mathrm{d}t\nonumber\\
&\quad + \sum_{l=1}^{N}\Big({\bf L}_l\boldsymbol{\rho}_{t}^{N} + \boldsymbol{\rho}_{t}^{N}{\bf L}^{\dagger}_l -\tr\big(({\bf L}_l + {\bf L}_l^{\dagger})\boldsymbol{\rho}_{t}^{N}\big)\boldsymbol{\rho}_{t}^{N}\Big)\mathrm{d}W_t^{l}
\end{align}} 
Here $(\boldsymbol{\rho}_{t}^{N})_{t \geq 0}$ is a stochastic process under the filtered probability space $\bigl(\Omega,\mathcal{F},(\mathcal{F}_t)_{t \ge 0},\mathbb{P}\bigr)$, with $\bigl((W_t^{l})_{t \ge 0}\bigr)_{1 \le l \le N}$, a family of $N$ independent Wiener processes under the probability measure $\mathbb{P}$.

\item \textbf{Un}normalize - Linear
{\small
\begin{align}\label{Nunormalize}
\mathrm{d}\boldsymbol{\varrho}_t^{N} &= -\mathrm{i}{[\mathbf{H}^{N},\boldsymbol{\varrho}_{t}^{N}]}\,\mathrm{d}t
+ \sum_{l=1}^{N} \Big({\bf L}_l\,\boldsymbol{\varrho}_{t}^{N}\,{\bf L}^{\dagger}_{l} \;-\; \tfrac{1}{2}\big\{{\bf L}^{\dagger}_{l}\,{\bf L}_{l},\,\boldsymbol{\varrho}_{t}^{N}\big\}\Big)\,\mathrm{d}t\nonumber\\
&\quad + \sum_{l=1}^{N}\Big({\bf L}_l\,\boldsymbol{\varrho}_{t}^{N} \;+\; \boldsymbol{\varrho}_{t}^{N}\,{\bf L}^{\dagger}_l\Big)\,\mathrm{d}Y_t^{l}
\end{align}
}
Here $(\boldsymbol{\varrho}_{t}^{N})_{t \geq 0}$ is a stochastic process under the filtered probability space $\bigl(\Omega,\mathcal{F},(\mathcal{F}_t)_{t \ge 0},\mathbb{Q}\bigr)$, with $\bigl((Y_t^{l})_{t \ge 0}\bigr)_{1 \le l \le N}$, a family of $N$ independent Wiener processes under the probability measure $\mathbb{Q}$.\footnote{Note that the process $\boldsymbol{\varrho}_{t}^{N}$ is not really a state, because $\tr(\boldsymbol{\varrho}_t^{N}) \neq 1 $.}
\end{itemize}
For the relation between \eqref{Nnormalize} and \eqref{Nunormalize} and more details, see \cite{bar09}.

In each case if we average over trajectories we get $N$-Body Lindblad equation i.e $$\mathbb{E}_{\mathbb{P}}[\boldsymbol{\rho}_{t}^{N}] = \boldsymbol{\bar{\rho}}_t^{N} = \mathbb{E}_{\mathbb{Q}}[\boldsymbol{\varrho}_t^{N}].
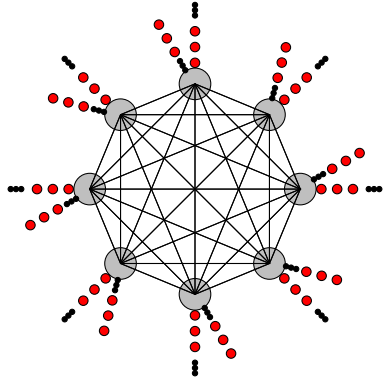
$$

\begin{center}
\begin{tikzpicture}[scale = 0.7]
  \foreach \x in {0,45,90,135,180,225,270,315}
  {
    \draw[fill = lightgray] (\x:2) circle (0.3);
    \draw[fill = red] (\x:2) ++(\x:0.4) circle (0.09);
    \draw[fill = red] (\x:2) ++(\x:0.7) circle (0.09);
    \draw[fill = red] (\x:2) ++(\x:1.0) circle (0.09);

    \draw[fill = black] (\x+1:2) ++(\x+30:0.3) circle (0.05);
    \draw[fill = black] (\x+1:2) ++(\x+30:0.4) circle (0.05);
    \draw[fill = black] (\x+1:2) ++(\x+30:0.5) circle (0.05);

    \draw[fill = black] (\x:2) ++(\x:1.3) circle (0.05);
    \draw[fill = black] (\x:2) ++(\x:1.4) circle (0.05);
    \draw[fill = black] (\x:2) ++(\x:1.5) circle (0.05);

    \draw[fill = red] (\x+1:2) ++(\x+30:0.7) circle (0.09);
    \draw[fill = red] (\x+1:2) ++(\x+30:1.0) circle (0.09);
    \draw[fill = red] (\x+1:2) ++(\x+30:1.3) circle (0.09);
  }


  \foreach \x in {0,45,90,135,180,225,270,315}
    \foreach \y in {0,45,90,135,180,225,270,315}
      \draw (\x:2) -- (\y:2);

\end{tikzpicture}
\begin{center}
\textbf{Schematic picture :} $N$-interacting quantum particles (Gray) subject to environement (Ancillas in red).  
\end{center}
\end{center}

\subsubsection*{Mean-field approximation}
From \cite{kolokoltsov21law,kolokoltsov22qmfg} one can derive rigorously the Mean-field quantum trajectories
\begin{itemize}
\item Normalize - \textbf{Non}-Linear
\begin{align*}
\mathrm{d}\gamma_{t} &= -\mathrm{i}[H + A^{m_t},\gamma_{t}]\mathrm{d}t + (L\gamma_{t}L^{\dagger} - \frac{1}{2}\{L^{\dagger}L,\gamma_{t}\})\mathrm{d}t+ (L\gamma_{t} + \gamma_{t}L^{\dag} - \tr((L+L^{\dagger})\gamma_t)\gamma_{t})\mathrm{d}W_t
\end{align*}

\item \textbf{Un}normalize - \textbf{Non}-Linear\footnote{When we pass in the mean-field limit we lose linearity in the drift part, but the diffusive remains linear for unormalize equation.}
\begin{align*}
\mathrm{d}\vartheta_{t} &= -\mathrm{i}[H + A^{m_t},\vartheta_{t}]\mathrm{d}t + (L\vartheta_{t}L^{\dagger} - \frac{1}{2}\{L^{\dagger}L,\vartheta_{t}\})\mathrm{d}t+ \big(L\vartheta_{t} + \vartheta_{t}L^{\dagger}\big)\mathrm{d}Y_t
\end{align*}

\item The mean-field operator \small{$A^{\bullet} : L^{2}_{\mathbb{C}}(\mathfrak{X}^2) \to L^{2}_{\mathbb{C}}(\mathfrak{X}^2) $}
\small{\begin{align*}
 A^{m}(x,y) &= \int_{\mathfrak{X}^2}a(x,y;x',y')\overline{m(x',y')}\mu(\mathrm{d}x'\mathrm{d}y').
\end{align*}}
\end{itemize}

In same manner averaging over trajectories in each case gets the Non-linear Lindblad equation\footnote{We can also call the equation mean-field Lindblad equation or Hartree-Lindblad equation.} i.e $\mathbb{E}_{\mathbb{P}}[{\gamma}_{t}] = m_t = \mathbb{E}_{\mathbb{Q}}[\vartheta_t]$. Where $(m_t)_{t \geq 0}$ is solution of the following non-linear ODE :
$$ \frac{\mathrm{d}m_t}{\mathrm{d}t} = -\mathrm{i}[H + A^{m_t},m_t] + Lm_tL^{\dagger} - \frac{1}{2}\{L^{\dagger}L,m_t\}.$$

The justification of the mean-field limit is done through the notion of propagation of chaos. The idea originated from Kac \cite{kac56} for classical particles system. The quantum analogue was first introduced by Spohn in his famous paper \cite{spohn80kinetic}.

Basically the statement is that if the system is initially in a chaotic (i.e., tensor-product) state, it remains so at later times. i.e
\begin{align*}
    \boldsymbol{\rho}_{0}^{N} = \rho_{0}^{\otimes N} \xrightarrow[N \to \infty]{\text{Propagation of chaos}} \boldsymbol{\rho}_{t}^{N}  \approx \gamma_{t}^{\otimes N}. 
\end{align*}

Because quantum trajectories are inherently stochastic, this approximation must be understood in law. Taking expectations under the probability measure $\mathbb{P}$ yields the mean-field approximation for the Lindblad dynamics:  
\begin{align*}
    \mathbb{E}_{\mathbb{P}}[\boldsymbol{\rho}_t^{N}]
    &\approx \mathbb{E}_{\mathbb{P}}\big[\bigotimes_{l=1}^{N} \gamma_{t,l}\big]\\
    &\approx \prod_{l=1}^{N}\mathbb{E}_{\mathbb{P}}[\boldsymbol{\gamma}_{t,l}] = \mathbb{E}_{\mathbb{P}}[\gamma_t]^{\otimes N}
\end{align*}
which can also be written as,
\begin{align*}
    \boldsymbol{\bar{\rho}}_t^{N} &\approx m_t^{\otimes N}.
\end{align*}

Here $\gamma_{t,l}$ is an independent copy of $\gamma_{t}$ driven by an independent Wiener process.

\subsubsection*{Pure state trajectories}
Quantum trajectories preserve pure state, 

\begin{proposition}
If the initial state is pure state then it remains pure state at later times i.e :
$$\gamma_{0} = \ket{\psi_0}\bra{\psi_0} \xrightarrow[]{\text{Purity preservation}}  \gamma_t = \ket{\psi_t}\bra{\psi_t}$$ 
$$\vartheta_{0} = \ket{\chi_0}\bra{\chi_0} \xrightarrow[]{\text{Purity preservation}}  \vartheta_t = \ket{\chi_t}\bra{\chi_t}$$ 
\end{proposition}

As consequence we can write the dynamics in pure state, 

\begin{itemize}
\item Normalize - \textbf{Non}-Linear
{ \begin{align*}
\mathrm{d}\psi_{t} &= -\mathrm{i}\Bigg(H + A^{\mathbb{E}_{\mathbb{P}}[\ket{\psi_t}\bra{\psi_t}]}\Bigg)\psi_t\mathrm{d}t - \frac{1}{2}\Bigg(L^{\dagger}L - \langle L^{\dagger}L\rangle_{\psi_t}\Bigg)\psi_t\mathrm{d}t+ (L - \frac{1}{2}\langle L\rangle_{\psi_t})\psi_t\mathrm{d}W_t
\end{align*}}

\item \textbf{Un}normalize - \textbf{Non}-Linear
{ \begin{align*}
\mathrm{d}\chi_{t} &= -\mathrm{i}\Bigg(H + A^{\mathbb{E}_{\mathbb{Q}}[\ket{\chi_t}\bra{\chi_t}]} \Bigg)\chi_t\mathrm{d}t - \frac{1}{2}L^{\dagger}L\chi_t\mathrm{d}t + L\chi_t\mathrm{d}Y_t
\end{align*}}
\end{itemize}

\subsubsection*{Particles Methods}

For the sake of conciseness, we will detail the procedure only for the unraveling with $(\psi_t)_{t \geq 0}$. The case of $(\chi_t)_{t \geq 0}$ is handled similarly. 

The stochastic process $(\psi_t)_{t \geq 0}$ is issued from a stochastic differential equation of McKean-Vlasov type were the dependence in law is through the expectation in the drift part, one can approach the mean-field particle by a system of interacting particles : 
{\small \begin{align*}
\mathrm{d}\psi_{t,l} &= -\mathrm{i}\Bigg(H + A^{\frac{1}{N}\sum_{l'=1}^{N}\ket{\psi_{t,l'}}\bra{\psi_{t,l'}}}\Bigg)\psi_{t,l}\mathrm{d}t \\&-\frac{1}{2}\Bigg(L^{\dagger}L - \langle L^{\dagger}L\rangle_{\psi_{t,l}}\Bigg)\psi_{t,l}\mathrm{d}t+ (L - \frac{1}{2}\langle L\rangle_{\psi_{t,l}})\psi_{t,l}\mathrm{d}W_t^{l}.
\end{align*}}

Using linearity of $A^{\bullet}$ we can re-write the system in :
\begin{itemize}
\item Pure state form (Wave function) :
{\small \begin{align*}
\mathrm{d}\psi_{t,l} &= -\mathrm{i}\Bigg(H + \frac{1}{N}\sum_{l'=1}^{N}A^{\ket{\psi_{t,l'}}\bra{\psi_{t,l'}}}\Bigg)\psi_{t,l}\mathrm{d}t-\frac{1}{2}\Bigg(L^{\dagger}L - \langle L^{\dagger}L\rangle_{\psi_{t,l}}\Bigg)\psi_{t,l}\mathrm{d}t\\&+ (L - \frac{1}{2}\langle L\rangle_{\psi_{t,l}})\psi_{t,l}\mathrm{d}W_t^{l},
\end{align*}}
\item Density operator form :
{\small \begin{align*}
\mathrm{d}\gamma_{t,l} &= -\mathrm{i}[H + \frac{1}{N}\sum_{l'=1}^{N}A^{\gamma_{t,l'}},\gamma_{t,l}]\mathrm{d}t + (L\gamma_{t,l}L^{\dagger} - \frac{1}{2}\{L^{\dagger}L,\gamma_{t}\})\mathrm{d}t\\ &\quad+ \Bigg(L\gamma_{t,l} + \gamma_{t,l}L^{\dagger} - \tr((L+L^{\dagger})\gamma_{t,l})\gamma_{t,l}\Bigg)\mathrm{d}W_t^{l}.
\end{align*}}
\end{itemize}

So to simulate one mean-field quantum trajectory, we simulate $N$-interacting quantum trajectories. By averaging over all these trajectories, we approximate the nonlinear Lindblad equation i.e :
{\small\begin{align}\label{particle-approx}
    m_t &\approx  \frac{1}{N}\sum_{l=1}^{N}\gamma_{t,l}\\
    &\approx \frac{1}{N}\sum_{l=1}^{N}\ket{\psi_{t,l}}\bra{\psi_{t,l}}
\end{align}}

In the next section we provide a convergence result, that justify \eqref{particle-approx}.

\section*{Classical Propagation of Chaos}
\begin{lemma}[Theorem 1.4.8 \cite{khanfer24applied}]
If $a \in L^2_{\mathbb{C}}(\mathfrak{Z}\times\mathfrak{Z},\tilde{\mu}^{\otimes 2})$, where $(\mathfrak{Z},\Omega_{\mathfrak{Z}},\tilde{\mu})$ is measure space  then the integral operator $$A : L^{2}_{\mathbb{C}}(\mathfrak{Z},\tilde{\mu}) \to L^{2}_{\mathbb{C}}(\mathfrak{Z},\tilde{\mu}) $$  is Hilbert-Schmidt operator and $\|A\|_{\text{HS}} = \|a\|_{2} $
\end{lemma}

\begin{lemma}
Let $A$ mesurable complex-valued function on $\mathfrak{X}^2$ of Hilbert-Schmidt with kernel $a$ a complex-valued function on $\mathfrak{X}^4$  and $\rho$ complex-valued function on $\mathfrak{X}^2$, let 
$$ A^{\rho}(x,y) = \int_{\mathfrak{X}^2} a(x,y;x',y')\overline{\rho(x',y')}\mu(\mathrm{d}x'\mathrm{d}y')$$
Then, 
$$
\|A^{\rho}\|^2_{L^2(\mathfrak{X}^2)} \leq \|{\rho}\|^2_{L^2(\mathfrak{X}^2)} \|a\|^2_{L^2(\mathfrak{X}^4)}.
$$
\end{lemma}

\begin{theorem}[Monte-Carlo Approximation]\label{thm:main}
There exists a constants $C_1, C_2>0$ (independent of $N$) such that
{\small \begin{align*} 
&\text{1}-
\sup_{t \in [0,T]}
\mathbb{E}_{\mathbb{P}}\biggl[
  \tr\Bigg(\Big(m_t \;-\; \tfrac{1}{N}\sum_{l=1}^N \gamma_{t,l}\Big)^{2}
  \Bigg)
\biggr]
\leq
\frac{C_1}{N}.\\
&\text{2}-
\sup_{t \in [0,T]}
\mathbb{E}_{\mathbb{Q}}\biggl[
  \tr\Bigg(\Big(m_t \;-\; \tfrac{1}{N}\sum_{l=1}^N \vartheta_{t,l}\Big)^{2}
  \Bigg)
\biggr]
\leq
\frac{C_2}{N}.
\end{align*}}
\end{theorem}

The following corollary justifies the use of one trajectory of the particle system to represent the mean-field quantum trajectory.

\begin{corollary}[Classical Propagation of chaos]
  There exists a constants $C_3, C_4>0$ (independent of $N$) such that
{\small \begin{align*} 
&\text{1}-
\sup_{l \in [N]}\sup_{t \in [0,T]}
\mathbb{E}_{\mathbb{P}}\biggl[
  \Big\|\psi_t - \psi_{t,l}
  \Big\|_{2}^{2}
\biggr]
\;\le\;
\frac{C_3}{N}.\\
&\text{2}-
\sup_{l \in [N]}\sup_{t \in [0,T]}
\mathbb{E}_{\mathbb{Q}}\biggl[
  \Big\|\chi_t - \chi_{t,l}
  \Big\|_{2}^{2}
\biggr]
\;\le\;
\frac{C_4}{N}.
\end{align*}}  
\end{corollary}

\section*{Conclusion}
We have presented a mean-field particle Monte Carlo method for simulating the $N$-body Lindblad equation. We established the convergence (in law) of  interacting quantum particles to the nonlinear Lindblad equation in the large-$N$ limit. Future work could focus on refining the numerical discretization beyond the standard Euler scheme, for instance by re-adapting method proposed in \cite{rouchon15efficient}.

\section*{Acknowledgements}
This work was supported by the ANR projects Q-COAST (ANR-19-CE48-0003) and IGNITION (ANR-21- CE47-0015). This research was initiated during a visit to the {\it Institute for Mathematical and Statistical Innovation} (IMSI), which is supported by the National Science Foundation ({Grant No. DMS-1929348}), and was later completed during a visit to the {\it International Centre for Theoretical Sciences} (ICTS) as part of the 'Quantum Trajectories' program.

\newpage
\section*{Appendix}
\subsection*{How operators act?}
In order to implement a program that run  algorithm, we have to make the correspondence between the functional and matrix of an operator.

Consider $\mathfrak{X} = \{1,\dots,d\}$ with $\mu$ counting measure, then spaces $L^{2}_{\mathbb{C}}(\mathfrak{X},\mu)$,  reduce to finite-dimensional Hilbert spaces, and we have the following isomorphisms : 
$$ L^{2}_{\mathbb{C}}(\mathfrak{X},\mu) \simeq \mathbb{C}^d. $$

In same manner for $L^{2}_{\mathbb{C}}(\mathfrak{X}^2,\mu^{\otimes 2})$ and $L^{2}_{\mathbb{C}}(\mathfrak{X}^4,\mu^{\otimes 4})$
$$L^{2}_{\mathbb{C}}(\mathfrak{X}^2,\mu^{\otimes 2}) \simeq \mathbb{C}^{d^2} \simeq \mathcal{M}_{d}(\mathbb{C}), \quad  L^{2}_{\mathbb{C}}(\mathfrak{X}^4,\mu^{\otimes 4}) \simeq \mathbb{C}^{d^4} \simeq \mathcal{M}_{d^2}(\mathbb{C}).$$

\begin{remark}
The dynamics of an open quantum systems with infinite-dimensional state space can be approximated by finite dimensional  master equations. Indeed, if $\dim(\mathbb{H})= \infty$ we can always find a finite dimensional subspace $\mathbb{H}^{d}$ of $\mathbb{H}$ such that $H_d\approx H \approx P_dHP_d$ same with $L $ and $A$,  where $P_d$ is the orthogonal projection of $\mathbb{H}$ onto $\mathbb{H}^d$. 
\end{remark}

\noindent
Consider an operator $K$ function of two variable i.e an element in $L^{2}_{\mathbb{C}}(\mathfrak{X}\times\mathfrak{X},\mu^{\otimes 2})$, we can construct a correspondance of $K$ in  matrix formulation i.e element in $\mathcal{M}_{d}(\mathbb{C})$, 
{\small $$ K = \begin{pmatrix}
    K(1,1) & K(1,2) & \dots & K(1,d)\\
    K(2,1) & K(2,2) & \dots & K(2,d)\\
    \vdots & \vdots & \rotatebox{45}{\vdots} & \vdots\\
    K(d,1) & K(d,2) & \dots & K(d,d)
\end{pmatrix} \in \mathcal{M}_{d}(\mathbb{C})
$$}

We can make correspondance between $\mathcal{M}_{d}(\mathbb{C}) $ and $\mathbb{C}^{d^2} $ by taking vectorisation :
\begin{align*} K \in \mathcal{M}_{d}(\mathbb{C}) &\leftrightarrow  \textbf{vec}(K^{\dagger}) \in \mathbb{C}^{d^2}\\
\mathbf{vec}(K^{\dagger}) &= \begin{pmatrix} K(1,1) \\ \overline{K(1,2)} \\ \overline{K(1,3)} \\ \vdots \\ \overline{K(1,d)} \\ \overline{K(2,1)} \\ \vdots  \\ K(d,d) \end{pmatrix}
\end{align*}

\paragraph{Mean-field interaction operator}
Let consider the Hilbert space $\mathbb{H}\otimes\mathbb{H} = L^{2}_{\mathbb{C}}(\mathfrak{X},\mu)\otimes L^{2}_{\mathbb{C}}(\mathfrak{X},\mu) = L^{2}_{\mathbb{C}}(\mathfrak{X}^2,\mu^{\otimes 2}) $, and a Hilbert-Schmidt operator $A$ acting on $\mathbb{H}\otimes\mathbb{H}$ with kernel $a$, with the following properties : 
\begin{align*}
    a(x,y;x',y') &= a(y,x;y',x'), \; a(x,y,x',y') = \overline{a(x',y';x,y)}
\end{align*}

The mean-field operator $A^{\bullet}$ act on density operataor $\gamma $ as follow :
\begin{itemize}
\item Function formulation : 
\begin{align*}
    A^{\gamma}(x,y) = \int_{\mathfrak{X}^{2}}a(x,y;x',y')\overline{\gamma(x',y')}\mu(\mathrm{d}x')\mu(\mathrm{d}y')
\end{align*}
\item Matrix formulation :
$$A^{\gamma} =  \textbf{devec}\big(a\textbf{vec}(\gamma^{\dagger})\big)$$
\end{itemize}

\paragraph{Example :}
In concrete example of qubit i.e $\mathbb{H}=\mathbb{C}^2$. Any density operator on $\mathcal{S}(\mathbb{C}^2)$ has the following Pauli decomposition :
$$\gamma_t = \frac{\mathbf{1} + x_t\sigma_{x} + y_t\sigma_{y} + z_t\sigma_{z}}{2}  = \frac{1}{2}\begin{pmatrix} 1 - z_t & x_t - \mathrm{i} y_t \\
    x_t + \mathrm{i} y_t & 1 + z_t \end{pmatrix}$$
$$ x_{t}^{2} + y_{t}^{2} + z_{t}^{2} \leq 1$$

Consider Hilbert-Schmidt operator $A$ with kernel $a$ define as follow  : 
\begin{align*}
a(1,1;1,1) &= \overline{a(2,2;2,2)} = 1, \quad
a(x,y,x',y') = 0 \; \text{otherwise.}
\end{align*}

In matrix formulation  :
$$a = \begin{pmatrix}
    1 & 0 & 0 & 0 \\
    0 & 0 & 0 & 0 \\
    0 & 0 & 0 & 0 \\
    0 & 0 & 0 & 1
\end{pmatrix}$$

Then the mean-field operator $A^{\bullet} $ is equal to :

\subparagraph{Matrix Formulation :}

\begin{align*}
A^{\gamma} &= \textbf{devec}[a \textbf{vec}[{m}^{\dagger}]]\\
&=  \begin{pmatrix}
    1 & 0 & 0 & 0 \\
    0 & 0 & 0 & 0 \\
    0 & 0 & 0 & 0 \\
    0 & 0 & 0 & 1
\end{pmatrix} \begin{pmatrix}
    {\gamma(1,1)}\\
    {\gamma(2,1)}\\
    {\gamma(1,2)}\\
    {\gamma(2,2)}
\end{pmatrix}\\
&= \textbf{devec}\Bigg[\begin{pmatrix}
    {\gamma(1,1)}\\
    0\\
0\\
{\gamma(2,2)}
\end{pmatrix}\Bigg] =   \frac{1}{2}\begin{pmatrix} 1 - z & 0 \\ 
0 & 1 + z
\end{pmatrix}
\end{align*},

$$A^{\gamma} = \frac{1}{2}\begin{pmatrix} 1 - z & 0 \\ 
0 & 1 + z 
\end{pmatrix} = \frac{1}{2}\Big(\mathbf{1} - z\sigma_z \Big)  $$

\subparagraph{Function Formulation :}

\begin{align*}
A^{\gamma}(x,y) = \int_{\mathfrak{X}^{2}}a(x,y;x',y')\overline{\gamma(x',y')}\mu(\mathrm{d}x')\mu(\mathrm{d}y') 
\end{align*}

Then, 
{\small \begin{align*}
A^{\gamma}(1,1) &= \sum_{(x,y) \in \mathfrak{X}^2} a(1,1;x,y)\overline{\gamma(x,y)}\\
&=a(1,1;1,1)\overline{\gamma(1,1)} + a(1,1;1,2)\overline{\gamma(1,2)} + a(1,1;2,1)\overline{\gamma(2,1)} + a(1,1;2,2)\overline{\gamma(2,2)}\\
&= \overline{m(1,1)}\\
&= \frac{1}{2}(1 - z),\\
A^{\gamma}(2,2) &= \frac{1}{2}(1 + z),\\
A^{\gamma}(1,2) &= A^{\gamma}(2,1) = 0.
\end{align*}}

$$ 
A^{\gamma} = \frac{1}{2}\begin{pmatrix} 1 - z & 0 \\ 
0 & 1 + z
\end{pmatrix}
$$

\subsection*{Algorithm}
Below is an Euler-type scheme for the interacting quantum trajectories approach.\footnote{See \cite{phdliu} for more details on McKean–Vlasov discretization.  In practice, step-size adaptivity or higher-order schemes may improve performance.}\\
\begin{algorithm}[H]
  \SetAlgoLined
  \KwIn{$(T,\rho_0,H,L,a)$}
  \KwOut{$m, A^m$}

  \textbf{Initialization:} $k \gets 0,\; m_{0} \gets \rho_{0}$\; 
  \textbf{Time step:} $\Delta t = \frac{T}{N}$\;

  \While{$k\,\Delta t \le T$}{
    \For{$l=1,\dots,N$}{
      $A^{\ket{\psi_{k,l}}\bra{\psi_{k,l}}} \gets 
      \text{devect}\Bigl[a\,*\,\text{vect}\bigl(\overline{\ket{\psi_{k,l}}\bra{\psi_{k,l}}}\bigr)\Bigr]$\;
      
      $\psi_{k+1,l} \gets 
      \psi_{k,l} 
      - \mathrm{i} \bigl( H + A^{m_{k}} \bigr)\psi_{k,l}\Delta t
      - \tfrac12 \Bigl( L^{\dagger}L - \langle L^{\dagger}L\rangle_{\psi_{k,l}}\Bigr)\,\psi_{k,l}\,\Delta t
      + \Bigl( L - \langle L\rangle_{\psi_{k,l}}\Bigr)\,\psi_{k,l}\sqrt{\Delta t}\mathcal{N}(0,1)$
    }
        $m_{k+1} \gets \frac{1}{N}\sum_{l=1}^{N}\ket{\psi_{k+1,l}}\bra{\psi_{k+1,l}}$\;
    $A^{m_{k+1}} \gets \frac{1}{N}\sum_{l=1}^{N} A^{\ket{\psi_{k+1,l}}\bra{\psi_{k+1,l}}}$\;
    $k \gets k+1$\;
  }
  \caption{Unraveling + Particle Method (Euler scheme)}
\end{algorithm}

\subsection*{Summary of the method}
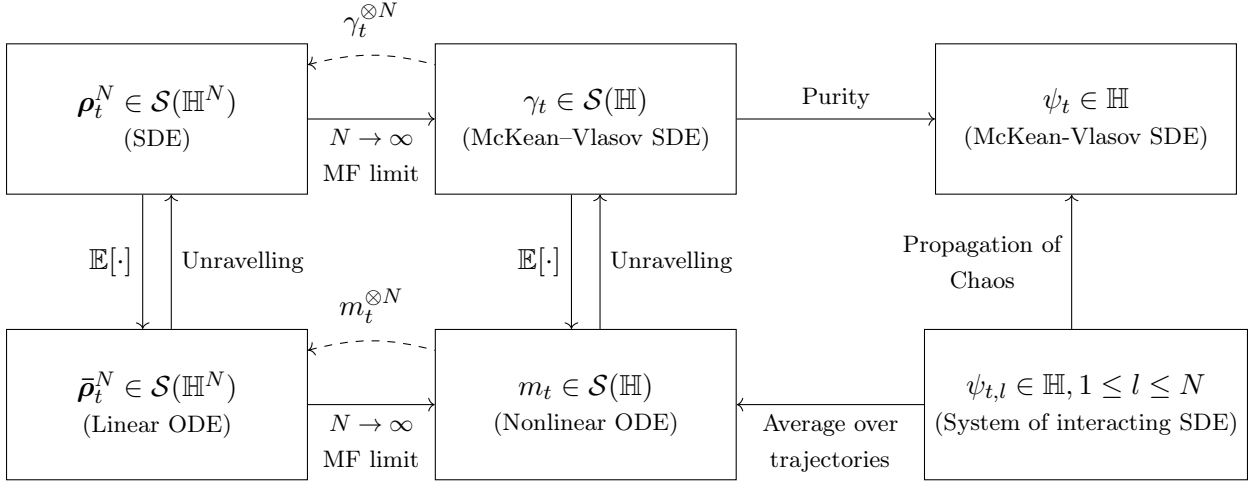
\begin{figure*}[th]
\centering
\begin{tikzpicture}[scale = 0.95][>=latex]

\tikzset{
  systembox/.style={
    draw,
    rectangle,
    minimum width=4cm,
    minimum height=2cm,
    align=center
  }
}

\node[systembox] (TL) at (0,0) {
  $\boldsymbol{\rho}_t^N \in \mathcal{S}(\mathbb{H}^N)$\\
  \footnotesize \text{(SDE)}
};

\node[systembox] (TR) at (6,0) {
  $\gamma_t \in \mathcal{S}(\mathbb{H})$\\
  \footnotesize \text{(McKean--Vlasov SDE)}
};

\node[systembox] (BL) at (0,-4) {
  $\boldsymbol{\bar{\rho}}_t^N \in \mathcal{S}(\mathbb{H}^N)$\\
  \footnotesize \text{(Linear ODE)}
};

\node[systembox] (BR) at (6,-4) {
  $m_t \in \mathcal{S}(\mathbb{H})$\\
  \footnotesize \text{(Nonlinear ODE)}
};

\node[systembox] (TR2) at (13,0) {
  $\psi_t \in \mathbb{H}$\\
  \footnotesize \text{(McKean-Vlasov SDE)}
};

\node[systembox] (BR2) at (13,-4) {
  $\psi_{t,{l}} \in \mathbb{H}, 1 \leq l \leq N$\\
  \footnotesize \text{(System of interacting SDE)}
};

\draw[->] (TR) -- node[above]{\footnotesize Purity} (TR2);

\draw[<-]
  ($(TR2.south) + (-0.2,0)$)
  -- node[left,align=center]{\footnotesize Propagation of \\ \footnotesize Chaos}
  ($(BR2.north) + (-0.2,0)$);

\draw[->] (BR2) -- node[below,align=center]{\footnotesize Average over\\ \footnotesize trajectories} (BR);

\draw[->] (TL) -- node[below,align=center]{\footnotesize $N \to \infty$\\ \footnotesize MF limit} (TR);
\draw[->] (BL) -- node[below,align=center]{\footnotesize $N \to \infty$\\ \footnotesize MF limit} (BR);

\draw[->]
  ($(TL.south) + (-0.2,0)$)
  -- node[left,align=center]{$\mathbb{E}[\cdot]$}
  ($(BL.north) + (-0.2,0)$);

\draw[->]
  ($(BL.north) + (0.2,0)$)
  -- node[right,align=center]{\footnotesize Unravelling}
  ($(TL.south) + (0.2,0)$);

\draw[->]
  ($(TR.south) + (-0.2,0)$)
  -- node[left,align=center]{$\mathbb{E}[\cdot]$}
  ($(BR.north) + (-0.2,0)$);

\draw[->]
  ($(BR.north) + (0.2,0)$)
  -- node[right,align=center]{\footnotesize Unravelling}
  ($(TR.south) + (0.2,0)$);

\draw[<-, bend left=15, dashed]
  ($(TL.north east)+(0,-0.3)$)
  to node[above, yshift=0.1cm]{$\displaystyle \gamma_t^{\otimes N}$}
  ($(TR.north west)+(0,-0.3)$);

\draw[<-, bend left=15, dashed]
  ($(BL.north east)+(0,-0.3)$)
  to node[above, yshift=0.1cm]{$\displaystyle m_t^{\otimes N}$}
  ($(BR.north west)+(0,-0.3)$);

\end{tikzpicture}
\caption{Overview of the method}
\label{fig:overview}
\end{figure*}

\newpage

\subsection*{Technicals estimates and proof}
\paragraph{Lemma proof}
\begin{proof}
Applying twice Cauchy-Schawrz inequality gets,
\begin{align*}
    \Big|A^{{\rho}}(x,y)\Big|^2 &\leq \Bigg(\int_{\mathfrak{X}^2} \Big|a(x,y;x',y')\overline{{\rho}(x',y')}\Big|\mu(\mathrm{d}x')\mu(\mathrm{d}y')\Bigg)\Bigg(\int_{\mathfrak{X}^2} \Big|a(x,y;t',z')\overline{{\rho}(t',z')}\Big|\mu(\mathrm{d}t')\mu(\mathrm{d}z')\Bigg),\\
    &\leq \Bigg(\int_{\mathfrak{X}^2}\Big|a(x,y;x',y')\Big|^2\mu(\mathrm{d}x')\mu(\mathrm{d}y')\Bigg)\Bigg(\int_{\mathfrak{X}^2} \big|{\rho}(t',z')\big|^2\mu(\mathrm{d}t')\mu(\mathrm{d}z')\Bigg),\\
    &= \|{\rho}\|^2_{L^2(\mathrm{X}^2)}\int_{\mathfrak{X}^2}\Big|a(x,y;x',y')\Big|^2\mu(\mathrm{d}x')\mu(\mathrm{d}y').
\end{align*}
Then integrating with respect to $\mu(\mathrm{d}x), \mu(\mathrm{d}y)$ yields
\begin{align*}
\int_{\mathfrak{X}^{2}}\big|A^{{\rho}}(x,y)\big|^{2}\mu(\mathrm{d}x)\mu(\mathrm{d}y)&\leq \|{\rho}\|^2_{L^2(\mathrm{X}^2)}\underbrace{\int_{\mathfrak{X}^2}\int_{\mathfrak{X}^2}\Big|a(x,y;x',y')\Big|^2\mu(\mathrm{d}x')\mu(\mathrm{d}y')\mu(\mathrm{d}x)\mu(\mathrm{d}y)}_{\|a\|_{L^2(\mathfrak{X}^4)}}.
\end{align*}
\end{proof}

\paragraph{Theorem proof }
\begin{proof}
In order to get the estimate we proceed à la Grönwall, first set $ \overline{\gamma}_t = \frac{1}{N}\sum_{l=1}^{N}\gamma_{t,l}$ then by Itô formula :
{\small 
\begin{align*}
\mathrm{d}\tr\Big((m_t - \overline{\gamma}_{t,l} )^2 \Big)  &= 2\tr\Big((m_t - \overline{\gamma}_{t,l} )(\mathrm{d}m_t - \mathrm{d}\overline{\gamma}_{t,l} )\Big) + \tr\big((\mathrm{d}\overline{\gamma}_t)^2\big)\\
= -2\mathrm{i}&\tr\Big((m_t - \overline{\gamma}_t)([H,m_t - \overline{\gamma}_t]+ [A^{m_t},m_t] -[A^{\overline{\gamma}_t},\overline{\gamma}_t])\Big)\mathrm{d}t\\ 
+ 2&\tr\Big((m_t - \overline{\gamma}_t)\big(L(m_t - \overline{\gamma}_t)L^{\dagger} - \frac{1}{2}\{LL^{\dagger},m_t - \overline{\gamma}_t\}\big)\Big)\mathrm{d}t\\
+ \frac{1}{N^2}\sum_{l=1}^{N}&\tr\Big(\Big(((L{\gamma}_{t,l} + {\gamma}_{t,l}L^{\dagger} - \tr((L+L^{\dagger}){\gamma}_{t,l}){\gamma}_{t,l}\Big)^2\Big)\mathrm{d}t + \sum_{l=1}^{N}f({\gamma}_{t,l})\mathrm{d}{W}_t^{l}
\end{align*}}

Taking the expectation we can re-write the expression,
{\small \begin{align*}
    \frac{\mathrm{d}\mathbb{E}_{\mathbb{P}}\Big[\tr\Big((m_t-\overline{\gamma}_t)^2\Big)\Big]}{\mathrm{d}t} &= \mathbb{E}_{\mathbb{P}}\Bigg[-2\mathrm{i}\tr\Big((m_t - \overline{\gamma}_t)[A^{m_t - \overline{\gamma}_t},m_t]\Big)+ \frac{1}{N^2}\sum_{l=1}^{N}\tr\Big(\Big(((L{\gamma}_{t,l} + {\gamma}_{t,l}L^{\dagger} - \tr((L+L^{\dagger}){\gamma}_{t,l}){\gamma}_{t,l}\Big)^2\Big)\\
 &+ 2\tr\Big((m_t - \overline{\gamma}_t)L(m_t - \overline{\gamma}_t)L^{\dagger} - LL^{\dagger}(m_t - \overline{\gamma}_t)^2\Big)\Bigg]
\end{align*}}
Taking triangular inequality, 
{\small 
\begin{align*}
\Bigg|\frac{\mathrm{d}\mathbb{E}_{\mathbb{P}}\Big[\tr\Big((m_t-\overline{\gamma}_t)^2\Big)\Big]}{\mathrm{d}t}&\Bigg| \leq 4(\star) + 2(\star\star)_{1} + 2 (\star\star)_{2} + (\star\star\star),
\end{align*}}

where,
{\small \begin{align*}
    (\star) &= \mathbb{E}_{\mathbb{P}}\Bigg[\Big|\tr\big((m_t - \overline{\gamma}_t)A^{m_t - \overline{\gamma}_t}m_t\big)\Big|\Bigg],  \\ 
    (\star\star)_1 &= \mathbb{E}_{\mathbb{P}}\Bigg[\Big|\tr\big((m_t - \overline{\gamma}_t)L(m_t - \overline{\gamma}_t)L^{\dagger}\big)\Big|\Bigg],\\
    (\star\star)_2 &= \mathbb{E}_{\mathbb{P}}\Bigg[\Big|\tr\big((m_t - \overline{\gamma}_t)^2LL^{\dagger}\big)\Big|\Bigg], \\
    (\star\star\star) &= \frac{1}{N^2}\sum_{l=1}^{N}\mathbb{E}_{\mathbb{P}}\Bigg[\Big|\tr\Big(\Big(((L{\gamma}_{t,l} + {\gamma}_{t,l}L^{\dagger}- \tr((L+L^{\dagger}){\gamma}_{t,l}){\gamma}_{t,l}\Big)^2\Big)\Big|\Bigg].
\end{align*}}
Then, 
{\small\begin{align*}
(\star) &\stackrel{\text{C-S}}\leq \mathbb{E}_{\mathbb{P}}\Bigg[\sqrt{\tr\Big((m_t-\overline{\gamma}_t)^2\Big)\tr\Big((A^{m_t-\overline{\gamma}_t})^{2}m_t^{2}\Big)}\Bigg],\\
&\stackrel{\tr(m^2)\leq 1}\leq \mathbb{E}_{\mathbb{P}}\Bigg[\|m_t - \overline{\gamma}_t\|_{2}\sqrt{\|A^{m_t-\overline{\gamma}_t}\|_{L^{2}(\mathfrak{X}^2)}} \Bigg],\\
&\stackrel{\text{Lemma 2}}\leq \mathbb{E}_{\mathbb{P}}\Bigg[\|m_t-\overline{\gamma}_t\|_{2}\|m_t-\overline{\gamma}_t\|_{L^{2}(\mathfrak{X}^2)}\|a\|_{L^2(\mathfrak{X}^4)}\Bigg],\\
    &= \|a\|_{2}\mathbb{E}_{\mathbb{P}}\Bigg[\tr\Big((m_t - \overline{\gamma}_{t})^2\Big) \Bigg].
\end{align*}}
For $(\star\star)$ we apply Cauchy-Schwarz directly, 
{\small \begin{align*}
    (\star\star)_{1} &\leq \|L\|_{2}^{2}\mathbb{E}_{\mathbb{P}}\Big[\tr\big((m_t-\overline{\gamma}_{t})^2\big)\Big],\\
    (\star\star)_{2} &\leq \|L\|_{2}^{2}\mathbb{E}_{\mathbb{P}}\Big[\tr\big((m_t-\overline{\gamma}_{t})^2\big)\Big],
\end{align*}}
same with $(\star\star\star)$,
{\small \begin{align*}
    (\star\star\star) &\leq \frac{2}{N^2}\sum_{l=1}^{N}\mathbb{E}_{\mathbb{P}}\Bigg[\tr\Big((L\gamma_{t,l}+\gamma_{t,l}L^{\dagger})^2\Big)+ \tr\Bigg(\Big(\tr\big((L+L^{\dagger})\gamma_{t,l}\big)\gamma_{t,l}\Big)^{2}\Bigg) \Bigg],\\
    &\leq \frac{2}{N^2}\sum_{l=1}^{N}\mathbb{E}_{\mathbb{P}}\Bigg[4\tr\Big((L\gamma_{t,l})^2\Big) + 2\tr^2\Big(L\gamma_{t,l}\Big) \Bigg],\\
    &\leq \frac{12\|L\|_{2}^{2}}{N}.
\end{align*}}
Putting all estimates together and set {\small$$\tilde{C}_1 = \max\big(4\|a\|_{2}+4\|L\|_{2}^{2}, 12\|L\|_{2}^{2}\big),$$}
we have
{\small \begin{align*}
    \Bigg|\frac{\mathrm{d}\mathbb{E}_{\mathbb{P}}\Big[\tr\Big((m_t-\overline{\gamma}_t)^2\Big)\Big]}{\mathrm{d}t}\Bigg| &\leq \tilde{C}_{1}\Bigg(\mathbb{E}_{\mathbb{P}}\Bigg[\tr\Big((m_t - \overline{\gamma}_t)^2\Big)\Bigg] + \frac{1}{N}\Bigg).
\end{align*}}

Then conclude with Grönwall lemma.

The proof is similar for the item $2$.

\end{proof}

\paragraph{Corollary proof}
\begin{proof}
In same manner to get the estimate we proceed à la Grönwall, fix $l \in \{1,\dots,N\}$ by Itô formula : 
\begin{align*}
\mathrm{d}\mathbb{E}_{\mathbb{Q}}\!\left[\|\chi_t-\chi_{t,l}\|_{2}^{2}\right]
&=
2\mathfrak{R}\Bigg(
\mathbb{E}_{\mathbb{Q}}\Big[
\langle \chi_t-\chi_{t,l},-\mathrm{i}H(\chi_t-\chi_{t,l})\rangle
+
\langle \chi_t-\chi_{t,l},-\mathrm{i}A^{m_t}(\chi_t-\chi_{t,l})\rangle\\
&\quad + \langle \chi_t-\chi_{t,l},\frac{1}{2}L^{\dagger}L(\chi_t-\chi_{t,l})\rangle +\big\langle \chi_t-\chi_{t,l},
-\mathrm{i}\!\Bigl(
A^{\frac{1}{N}\sum_{k=1}^{N}\ket{\chi_{t,k}}\bra{\chi_{t,k}}}
      - A^{m_t}\Bigr)\chi_{t,l}\big\rangle\Big]\Bigg)\mathrm{d}t\\ 
\quad &+ \mathbb{E}_{\mathbb{Q}}\!\left[
\langle \chi_t-\chi_{t,l},L^{\dagger}L(\chi_t-\chi_{t,l})\rangle
\right]\mathrm{d}t\\
&= \mathfrak{R}\Bigg(
\mathbb{E}_{\mathbb{Q}}\Big[
\langle \chi_t-\chi_{t,l},2\underbrace{\Big(-\mathrm{i}H - \mathrm{i}A^{m_t} + 2L^{\dagger}L\Big)}_{ := R}(\chi_t-\chi_{t,l})\rangle\Big]\Bigg)\mathrm{d}t\\
&\quad + \mathfrak{R}\Bigg(\mathbb{E}_{\mathbb{Q}}\Big[ \langle \chi_t-\chi_{t,l}, -\mathrm{i}A^{ \frac{1}{N}\sum_{k=1}^{N}\ket{\chi_{t,k}}\bra{\chi_{t,k}} - m_t }\chi_{t,l}\rangle\Big]\Bigg)\mathrm{d}t
\end{align*}
Applying Cauchy-Schwarz, 
\begin{align*}
    &\leq \|R\|_{2}\mathbb{E}_{\mathbb{Q}}\Big[\|\chi_{t} - \chi_{t,l}\|_{2}^{2}\Big] + \mathbb{E}_{\mathbb{Q}}\Big[\|A^{\frac{1}{N}\sum_{k=1}^{N}\vartheta_{t,l} - m_t }\|_{2}\|\chi_{t} - \chi_{t,l}\|_{2}\|\chi_{t,l}\|_{2} \Big],
\end{align*}
using the previous lemma, 
\begin{align*}
    &\leq \|R\|_{2}\mathbb{E}_{\mathbb{Q}}\Big[\|\chi_{t} - \chi_{t,l}\|_{2}^{2}\Big] + \mathbb{E}_{\mathbb{Q}}\Big[\|A\|_{2}\Big\|{\frac{1}{N}\sum_{k=1}^{N}\vartheta_{t,l} - m_t }\Big\|_{2}\|\chi_{t} - \chi_{t,l}\|_{2}\|\chi_{t,l}\|_{2} \Big],
\end{align*}
then thanks to estimate from previous theorem,
\begin{align*}
    &\leq \|R\|_{2}\mathbb{E}_{\mathbb{Q}}\Big[\|\chi_{t} - \chi_{t,l}\|_{2}^{2}\Big] + \frac{2C_{2}\|A\|_{2}}{N}.
\end{align*}
Put $\tilde{C}_{4} := \max\Big(\|R\|_{2},2C_{2}\|A\|_{2} \Big) $,

we have 
\begin{align*}
\Bigg|\frac{\mathrm{d}\mathbb{E}_{\mathbb{Q}}\Big[\Big\|\chi_{t} - \chi_{t,l} \Big\|_{2}^{2} \Big]}{\mathrm{d}t} \Bigg| &\leq \tilde{C}_{4}\Bigg( \mathbb{E}_{\mathbb{Q}}\Big[\Big\|\chi_{t} - \chi_{t,l} \Big\|_{2}^{2} \Big] + \frac{1}{N} \Bigg)  
\end{align*}

Then conclude with Grönwall lemma. Of course the proof is similar for the item $1$.

\end{proof}

{\footnotesize
\bibliographystyle{plain} 
\bibliography{Bibliography.bib}
}

\end{document}